\RequirePackage{ifpdf}
\ifpdf 
\documentclass[pdftex]{sigma}
\else
\documentclass{sigma}
\fi

\def\bR{\mathbb{R}}
\def\bC{\mathbb{C}}
\def\half{\frac{1}{2}}
\def\Tr{{\rm Tr }}
\def\A{\mathfrak{A}}

\begin{document}

\allowdisplaybreaks

\renewcommand{\thefootnote}{$\star$}

\renewcommand{\PaperNumber}{104}

\FirstPageHeading

\ShortArticleName{The Space of Connections as the Arena for (Quantum) Gravity}

\ArticleName{The Space of Connections as the Arena\\ for (Quantum) Gravity\footnote{This
paper is a contribution to the Special Issue ``Loop Quantum Gravity and Cosmology''. The full collection is available at \href{http://www.emis.de/journals/SIGMA/LQGC.html}{http://www.emis.de/journals/SIGMA/LQGC.html}}}

\Author{Steffen GIELEN}

\AuthorNameForHeading{S.~Gielen}

\Address{Albert Einstein Institute, Am M\"uhlenberg 1, 14476 Golm, Germany}
\Email{\href{mailto:gielen@aei.mpg.de}{gielen@aei.mpg.de}}

\ArticleDates{Received August 30, 2011, in f\/inal form November 09, 2011;  Published online November 11, 2011}

\Abstract{We review some properties of the space of connections as the natural arena for canonical (quantum) gravity, and compare to the case of the superspace of 3-metrics. We detail how a 1-parameter family of metrics on the space of connections arises from the canonical analysis for general relativity which has a natural interpretation in terms of invariant tensors on the algebra of the gauge group. We also review the description of canonical GR as a geodesic principle on the space of connections, and comment on the existence of a time variable which could be used in the interpretation of the quantum theory.}


\Classification{22E70; 51P05; 53C05; 53C80; 83C05; 83C45}

\renewcommand{\thefootnote}{\arabic{footnote}}
\setcounter{footnote}{0}

\vspace{-2mm}

\section{Introduction}

\looseness=-1
Loop quantum gravity (LQG) is a canonical quantisation of general relativity in connection va\-riables and the natural continuation of the geometrodynamics programme initiated by Wheeler, DeWitt and others in the 1950s for general relativity in the more conventional metric forma\-lism. As was realised early on in the study of geometrodynamics (see e.g.\ the review \cite{giulinisuper}), in order to understand canonical quantum gravity one must f\/irst understand the structure of the conf\/iguration space of the theory, famously denoted by Wheeler as {\it superspace}; it is an inf\/inite-dimensional manifold whose points correspond to (equivalence classes of) metrics on a 3-dimensional spatial slice $\Sigma$. The f\/irst systematic study of the properties of this space is probably due to DeWitt~\cite{dewitt}, and today the mathematical properties of Wheeler's superspace are fairly well understood. In contrast, the space of {\it connections} seems to have mainly been of interest in the study of Yang--Mills theory, and attracted less interest in the quantum gra\-vi\-ty literature, particularly for non-compact gauge group $G$ (e.g.\ $G=SO(3,1)$; we will mainly discuss four spacetime dimensions). Here, inspired by the  review \cite{giulinisuper} of the superspace of geo\-metrodynamics, we summarise some properties of the space of connections which are of interest in canonical quantum gravity, highlighting similarities and dif\/ferences to the space of metrics. After some def\/initions, we review the Hamiltonian analysis of GR in
connection variables for general Barbero--Immirzi parameter~$\gamma$
(for a previous derivation of the canonical formalism from
a different action see~\cite{desish}). The Hamiltonian constraint of the theory def\/ines a metric on the space of connections in the usual way. We argue that the one-parameter family of metrics def\/ined by all choices for $\gamma$ are precisely those local metrics linear in the curvature obtained by demanding gauge invariance; they correspond to all possible three-index invariant tensors for the adjoint representation of the gauge group $G$. We outline how pure GR can be understood to arise from a~geodesic principle on the space of connections. Finally, we comment on the existence of a~natu\-ral time variable on the space of connections and possible consequences on the interpretation of a~canonical quantisation of GR in connection variables, such as loop quantum gravity.

\section{The space of connections}

The central object in the following discussion will be a connection $\Omega$ on a principal f\/ibre bundle~$E$ which in physics is usually thought of as trivial, $E\simeq \Sigma\times G$, where $\Sigma$ is a 3-dimensional manifold thought of as a ``constant time slice'' in a 4-dimensional spacetime, and $G$ is the gauge group of the theory. Fixing a particular section in $E$, we will think of $\Omega$ as a one-form on $\Sigma$ valued in the Lie algebra $\mathfrak{g}$. Then the space of all such one-forms $\mathfrak{A}$ admits an action of the group $\mathfrak{G}$ of gauge transformations in $\Sigma$ corresponding to the change of section; elements of $\mathfrak{G}$ can be thought of as functions $G\rightarrow\Sigma$. In coordinates on $\Sigma$, $\mathfrak{G}$ acts as
\begin{gather*}
\Omega\mapsto g \Omega ,\qquad(g\circ \Omega)(x):=g^{-1}(x)\Omega(x)g(x)+g^{-1}(x)dg(x).
\end{gather*}
Of particular interest is then the space $\mathfrak{A}/\mathfrak{G}$ of connections up to gauge transformations, or an extension of it: In loop quantum gravity the space $\mathfrak{A}$ is extended to the space $\overline{\mathfrak{A}}$ of gene\-ra\-li\-sed connections, which can be def\/ined as the space of homomorphisms from the hoop group (piecewise analytic loops with f\/ixed base point) to the gauge group $G$. Clearly any connection~$\Omega$ def\/ines such a homomorphism through its holonomies, but $\overline{\mathfrak{A}}$ is larger in that it assumes no ``continuity'' on the homomorphisms which are left arbitrary.

If the gauge group $G$ is taken to be compact, one can then def\/ine the {\it Ashtekar--Lewandowski measure} \cite{ashlew} $d\mu_{{\rm AL}}(\Omega)$ on $\overline{\mathfrak{A}}$ through the normalised Haar measure on the group; f\/irst consider a~functional $f[\Omega]$ on $\overline{\mathfrak{A}}$ depending on a f\/inite number of (generalised) holonomies
\begin{gather*}
f[\Omega]\equiv f\left(H_{\gamma_1}(\Omega),\ldots,H_{\gamma_n}(\Omega)\right)
\end{gather*}
($\gamma_1,\ldots,\gamma_n$ are paths in $\Sigma$), which can be integrated with respect to $d\mu_{{\rm AL}}(\Omega)$ by
\begin{gather}
\int d\mu_{{\rm AL}}(\Omega)\,f[\Omega]\equiv \int dg_1\ldots dg_n\, f(g_1,\ldots,g_n).
\label{al}
\end{gather}
The measure is then rigorously def\/ined for all of $\overline{\mathfrak{A}}$ through projective limits, see~\cite{ashlew} for details. It is invariant under gauge transformations and 3-dif\/feomorphisms. Note that this is a purely group-theoretic construction without reference to the Hamiltonian constraint of any gravity theory. In Section~\ref{geomconn} we will encounter a metric on $\mathfrak{A}$, def\/ined by the Hamiltonian constraint of GR, which is also gauge invariant (so it projects to $\mathfrak{A}/\mathfrak{G}$) and (potentially) 3-dif\/feomorphism invariant; in that section we will work on a more formal level and not be able to construct a~rigorous def\/inition of a measure analogous to that induced by~(\ref{al}).

\section{Hamiltonian GR in connection variables}

It is well known that general relativity in four dimensions, with vanishing cosmological constant, can be def\/ined in terms of the action\footnote{Here and in the following, indices $a,b,c,\ldots$ denote internal indices for the Lorentz group $G$, $I, J, K, \ldots$ are internal indices of the compact subgroup of rotations, whereas $i,j,k,\ldots=1,2,3$ are spatial coordinate indices.}
\begin{gather}
S=\frac{1}{8\pi G}\int_{\Sigma\times\bR}\left(\half\epsilon_{abcd} e^a\wedge e^b\wedge R^{cd}[\omega]\right) ,
\label{einsthilb1}
\end{gather}
where spacetime is topologically of the form $\Sigma\times\bR$ for an unspecif\/ied 3-manifold topology $\Sigma$, $\omega^{ab}$ is a $\mathfrak{g}$-valued one-form viewed as a connection (the gauge group $G$ is $SO(3,1)$ or $SO(4)$, or an appropriate cover), $R^{ab}$ its curvature, and $e^a$ is an $\bR^4$-valued one-form representing an orthonormal frame. The following slight generalisation of (\ref{einsthilb1}) is the starting point for loop quantum gravity (LQG):
\begin{gather}
S=\frac{1}{8\pi G}\int_{\Sigma\times\bR}\left(\half\epsilon_{abcd} e^a\wedge e^b\wedge R^{cd}[\omega]+\frac{1}{\gamma}e^a\wedge e^b\wedge R_{ab}[\omega]\right) ,
\label{einsthilb2}
\end{gather}
where $\gamma$ is a real parameter, known as the Barbero--Immirzi parameter, which is fundamental in LQG. It was shown by Holst \cite{holst} that the canonical analysis of~(\ref{einsthilb2}) leads to the structure of LQG but the equivalent term in metric variables $\epsilon_{\mu\nu\rho\sigma}R^{\mu\nu\rho\sigma}$ in a gravity Lagrangian had been considered much earlier~\cite{hms}. Variation of (\ref{einsthilb2}) with respect to $e^a$ gives the equation of motion
\begin{gather}
\half\epsilon_{abcd}\big(e^b\wedge R^{cd}\big)+\frac{1}{\gamma}e^b\wedge R_{ab}=0\quad\rightarrow \quad \left({R^c}_a-\half R {\delta^c}_a\right)+\frac{s}{2\gamma}R_{adef}\epsilon^{defc}=0 ,
\label{einsteq}
\end{gather}
where we have written out the equation in components in the frame def\/ined by $e^a$ and $s=\pm 1$ is the spacetime signature. One recognises the vacuum Einstein equations plus a piece that will vanish if torsion is zero. The second equation of motion resulting from varying $\omega^{ab}$ is
\begin{gather}
\half\epsilon_{abcd}\big({-}T^a\wedge e^b+e^a\wedge T^b\big)+\frac{1}{\gamma}\left(-T_c\wedge e_d+e_c\wedge T_d\right)=0 ,
\label{torseq}
\end{gather}
where Cartan's f\/irst equation of structure $de^a=T^a-{\omega^a}_b\wedge e^b$ was used and it was then seen that all terms involving the connection cancel. (\ref{torseq}) implies that $T^a\wedge e^b=0$, and subsequently that the torsion two-form $T^a$ vanishes identically, if the two terms in (\ref{torseq}) are not Hodge dual to another, i.e.\ $\gamma^2\neq s$ (so in particular, for Lorentzian signature any real $\gamma$ is admissible). Then~(\ref{einsteq}) and~(\ref{torseq}) are equivalent to the equations of~GR.

Now focussing on the Hamiltonian formulation of (\ref{einsthilb2}), we perform the (3+1) splitting and def\/ine two one-forms and two 0-forms in $\Sigma$ by
\begin{gather*}
E^a\equiv e^a_i\,dx^i ,\qquad \chi^a\equiv e^a_t ,\qquad \Omega^{ab}\equiv\omega^{ab}_i\,dx^i ,\qquad \Xi^{ab}\equiv\omega^{ab}_t ,
\end{gather*}
where $i=1,2,3$ is now a spatial index and $t$ labels the time component (after we have f\/ixed a~time coordinate in the original spacetime $\Sigma\times\bR$). The action (\ref{einsthilb2}) then becomes
\begin{gather}
S   =   \frac{1}{8\pi G}\int_{\bR}\!dt\int_{\Sigma}\dot{\Omega}^{ab}\wedge\left(\half\epsilon_{abcd}E^c\wedge E^d+\frac{1}{\gamma}E_a\wedge E_b\right)+\chi^a\left(\epsilon_{abcd} E^b\wedge R^{cd}+\frac{2}{\gamma}E^b\wedge R_{ab}\right)\nonumber
\\
\phantom{S=}{} +  \Xi^{ab}\left(\half\epsilon_{abcd}D^{(\Omega)}\big(E^c\wedge E^d\big)+\frac{1}{\gamma}D^{(\Omega)}\left(E_a\wedge E_b\right)\right) ,
\label{hamact}
\end{gather}
where we have integrated by parts ignoring boundary terms (assuming $\Sigma$ to be compact) and the covariant exterior derivative $D^{(\Omega)}$ acts as
\begin{gather}
D^{(\Omega)}\big(E^a\wedge E^b\big)=d\big(E^a\wedge E^b\big)+{\Omega^a}_c\wedge E^c\wedge E^b + {\Omega^b}_c\wedge E^a\wedge E^c .
\label{covder}
\end{gather}
Determining the momentum conjugate to the connection $\Omega^{ab}$, one f\/inds the vector density
\begin{gather}
\pi^i_{ab}=\frac{1}{16\pi G}\epsilon^{ijk}\epsilon_{abcd}E^c_j E^d_k+\frac{1}{8\pi\gamma G}\epsilon^{ijk}E_{ja}E_{kb} ,
\label{momenta}
\end{gather}
while the momenta conjugate to $\Xi^{ab}$, $\chi^a$ and $E^a_i$ are seen to vanish. In Hamiltonian language, all these conditions give primary constraints which should be added to the original ``naive'' Hamiltonian. Their consistency under time evolution leads to secondary constraints, which split into two groups: Those obtained from conservation of (\ref{momenta}) and of the vanishing of the momenta~$P_a^i$ conjugate to~$E^a_i$ can be solved for some of the Lagrange multipliers; the other constraints $\mathcal{G}_{ab}=\frac{\delta S}{\delta\Xi^{ab}}$ and $\mathcal{H}_a=\frac{\delta S}{\delta\chi^{a}}$ are constraints on the dynamical variables.

Altogether one has a phase space parametrised by 40 variables $(\Omega^{ab}_i,\Xi^{ab},E^a_i,\chi^a)$ plus their canonical momenta for each point in $\Sigma$, subject to primary and secondary constraints. It is a~convenient procedure to remove $\chi^a$ and $\Xi^{ab}$ from the phase space and view them as Lagrange multipliers enforcing the constraints $\mathcal{G}_{ab}$ and $\mathcal{H}_a$ on the dynamical variables; it is also consistent to use (\ref{momenta}) to replace $E^a_i$ by $\pi^i_{ab}$ everywhere and to formulate all of the dynamics purely in terms of $(\Omega^{ab}_i,\pi^i_{ab})$ after adding six constraints enforcing $\pi^i_{ab}$ to be of the form (\ref{momenta})\footnote{If these constraints are not added to the Hamiltonian initally, they will be generated by consistency of the Hamiltonian constraint under time evolution, i.e.\ the Poisson brackets $\{\mathcal{H},\mathcal{H}\}$ \cite{barros}.}. These are the {\it simplicity constraints}
\begin{gather}
\mathcal{C}^{ij}\equiv \epsilon^{abcd}\Pi^i_{ab}\Pi^j_{cd} ,\qquad\Pi^i_{ab}\equiv\pi^i_{ab}-\frac{\gamma}{2}{\epsilon_{ab}}^{cd}\pi^i_{cd} ;
\label{simpl}
\end{gather}
consistency of those 
constraints under time evolution generates the additional constraints \cite{barros}
\begin{gather}
\mathcal{D}^{ij}\equiv \epsilon^{abcd}\Pi^k_{cd}\Pi^{(i}_{ae}D^{(\Omega)}_k{{\Pi^{j)}}_b}^e ,
\label{simpl2}
\end{gather}
where $D^{(\Omega)}_k$ is a (gauge-)covariant derivative as in (\ref{covder}). (\ref{simpl}) and (\ref{simpl2}) form a second class pair and therefore, by the 
Dirac algorithm, have to be solved before quantisation\footnote{Second class constraints also imply additional conditions on Lagrange multipliers, those used to enforce~$\mathcal{C}^{ij}$ and~$\mathcal{D}^{ij}$; if the Hamiltonian constraint itself would be second class, this would imply a condition on its corresponding Lagrange multiplier, the lapse, as seen for the ``$\lambda R$ model'' of Ho\v{r}ava--Lifshitz gravity in \cite{lambdar}.}. This is the motivation for passing to time gauge and reducing the gauge group from~$SO(3,1)$ to~$SO(3)$ or~$SU(2)$, as we will outline at the end of this section after having derived the f\/irst class constraints.

Among the constraints enforced by the Lagrange multipliers $\chi^a$, $\Xi^{ab}$ (i.e.\ the secondary constraints coming from vanishing of their conjugate momenta), we have the {\it Gauss constraint} familiar from gauge theories,
\begin{gather*}
\mathcal{G}_{ab}\equiv D_i^{(\Omega)}\pi^i_{ab} ,
\end{gather*}
which generates $G$ gauge transformations. The other four constraints split into dif\/feomorphism and Hamiltonian constraints, according to their interpretation in terms of gauge transformations (spatial/time dif\/feomorphisms). One now has (18+18) phase space variables per point in~$\Sigma$ and~10 f\/irst class and~12 second class constraints which reduce to two physical degrees of freedom.

To identify the dif\/feomorphism and Hamiltonian constraints, one can decompose the $\bR^4$-valued Lagrange multiplier $\chi^a$ as
\begin{gather*}
\chi^a=e^a_t=\frac{e^{ta}}{g^{tt}}-e^a_i \frac{e^t_b e^{ib}}{g^{tt}}\equiv -N^2\,e^{ta}+e^a_i N^i,
\end{gather*}
where we introduce the inverse tetrad $(e^\mu_a)$ and $g^{tt}\equiv e^t_a e^{ta}$; a nondegeneracy condition $\det e\neq 0$ has to be assumed as in other approaches to quantum gravity. It is rather unclear how to guarantee this at the quantum level\footnote{Wise \cite{wise} has suggested that the MacDowell--Mansouri formulation of gravity, where one unif\/ies vierbein and connection into an $\mathfrak{so}(4,1)$ connection, might shed light on this since $\det e\neq 0$ is the requirement for the $\mathfrak{so}(4,1)$ connection to be a Cartan connection.}. $N$ and $N^i$ are the usual lapse and shift of canonical~GR.

The term involving $\chi^a$ in (\ref{hamact}) can now be rewritten as{\samepage
\begin{gather}
\frac{\chi^a}{4\pi G}\left(\epsilon_{abcd}E^b_i \tilde{R}^{icd}+\frac{2}{\gamma}E^b_i\tilde{R}^i_{ab}\right)   =   -\frac{N^2}{4\pi G} e^{ta}\left(\epsilon_{abcd}E^b_i \tilde{R}^{icd}+\frac{2}{\gamma}E^b_i\tilde{R}^i_{ab}\right)
\label{const}
\\
\hphantom{\frac{\chi^a}{4\pi G}\left(\epsilon_{abcd}E^b_i \tilde{R}^{icd}+\frac{2}{\gamma}E^b_i\tilde{R}^i_{ab}\right)   =}{}
 +\frac{1}{8\pi G}\left(\epsilon_{abcd}\,E^a_l E^b_m\epsilon^{ilm}R^{cd}_{ij}N^j+\frac{2}{\gamma}E^a_l E^b_m\epsilon^{ilm}R_{ijab}N^j\right) ,\nonumber
\end{gather}
where we introduce a vector density $\tilde{R}^{iab}\equiv\half\epsilon^{ijk}R^{ab}_{jk}$ dual to the curvature 2-form.}

If we now introduce the scalar weight two densities
\begin{gather}
f(E,R)\equiv E^a_l E^c_m \epsilon^{ilm}R_{ijab}\epsilon^{jnp}E^b_n E_{pc},\qquad
g(E,R)\equiv E^d_l E^e_m \epsilon^{ilm}R_{ijab}\epsilon^{jnp}E^b_n E^c_p {\epsilon^a}_{cde} ,
\end{gather}
it is straightforward to f\/irst verify that
\begin{gather}
g(E,\star R)=2sf(E,R) ,\qquad (\star R)^{ab}\equiv\half{\epsilon^{ab}}_{cd}R^{cd} ,
\label{hodge}
\end{gather}
and so by linearity and $\star^2=s$ also $f(E,\star R)=\half g(E,R)$. Furthermore, one sees that
\begin{gather*}
g(E,R)=\frac{2}{3}\det\big(E_i^c,E_j^d,E_k^e\big){\epsilon^a}_{cde} E^b_i\tilde{R}^i_{ab}=4s (\det e) e^{ta} E^b_i\tilde{R}^i_{ab} ,
\end{gather*}
where we have used the invertibility of the matrix $(e_\mu^a)$ so that $(\det e)e_a^t=\frac{s}{6}\epsilon_{abcd}\det(E^b_i,E^c_j,E^d_k)$. With these def\/initions in hand, we rewrite (\ref{const}) as
\begin{gather*}
-\frac{N^2}{4\pi G(\det e)}\!\left(f(E,R)+\frac{s}{2\gamma}g(E,R)\right)+\frac{1}{8\pi G}\!\left[\epsilon_{abcd} E^a_l E^b_m\epsilon^{ilm}R^{cd}_{ij}+\frac{2}{\gamma}E^a_l E^b_m\epsilon^{ilm}R_{ijab}\right]N^j ;
\end{gather*}
to f\/inally express this in terms of the momenta $\pi^i_{ab}$ we need the following equalities,
\begin{gather*}
\pi^i_{ac}R^{ab}_{ij}{{\pi^j}_b}^c = \left(\frac{1}{8\pi G}\right)^2\left[\left(s+\frac{1}{\gamma^2}\right)f(E,R)+\frac{1}{\gamma}g(E,R)\right] ,
\\
\half{\epsilon^{ab}}_{ef}\pi^i_{ac}R^{ef}_{ij}{{\pi^j}_b}^c = \left(\frac{1}{8\pi G}\right)^2\left[\frac{2s}{\gamma}f(E,R)+\half\left(s+\frac{1}{\gamma^2}\right)g(E,R)\right] ,
\\\pi^i_{ab}R^{ab}_{ij} = \frac{1}{8\pi G}\left(\half\epsilon^{abcd}E^c_l E^d_m \epsilon^{ilm}R^{ab}_{ij}+\frac{1}{\gamma}E^a_l E^b_m \epsilon^{ilm}R_{ijab}\right) ,
\end{gather*}
which follow relatively straightforwardly from the def\/inition of $\pi^i_{ab}$. So (\ref{const}) can be written as
\begin{gather*}
\frac{2N^2}{\det e}\frac{8\pi\gamma^2 G}{(\gamma^2-s)}\left(\pi^i_{ac}\left(R^{ab}_{ij}-\frac{s}{2\gamma}{\epsilon^{ab}}_{ef}R^{ef}_{ij}\right){{\pi^j}_b}^c\right)+2 \pi^i_{ab}R^{ab}_{ij}N^j
\end{gather*}
and we have succeeded in identifying the dif\/feomorphism and Hamiltonian constraints
\begin{gather}
\mathcal{H}_j \equiv \pi^i_{ab}R^{ab}_{ij},\nonumber\\
\mathcal{H} \equiv \pi^i_{ac}\left(R^{ab}_{ij}-\frac{s}{2\gamma}{\epsilon^{ab}}_{ef}R^{ef}_{ij}\right){{\pi^j}_b}^c,
\label{konstf}
\end{gather}
which indeed satisfy the correct algebra as shown in~\cite{barros}.

Why would one consider $\gamma<\infty$ in (\ref{einsthilb2})? The parameter $\gamma$ does not modify the dynamical content of the theory, but the symplectic structure on phase space. In LQG the central object of interest is the {\em real Barbero connection} \cite{barbero} $A^{ab}=\Omega^{ab}+\frac{\gamma}{2}{\epsilon^{ab}}_{cd}\Omega^{cd}$ whose conjugate momentum is $\Pi^i_{ab}$ def\/ined in (\ref{simpl}). The one-form $A^{cd}$ can no longer be interpreted as a connection on a~$G$-bundle over $\Sigma$ (contrary to a claim made in \cite{barros}); however both its $\mathfrak{so}(3)$ part $A^{IJ}$ and the dualised complement $\frac{s}{\gamma}\epsilon^{IJK}A_{0K}$ can be viewed as connections on an $SO(3)$-bundle.

Lorentz symmetry is broken to $SU(2)$ by reducing the system to the hypersurface in phase space where the second class constraints are satisf\/ied, as done in~\cite{barros}. (The symmetry breaking can also be understood geometrically by considering local observers~\cite{derekgielen}.) The general solution to~(\ref{simpl}), assuming that all $\Pi^i_{IJ}$ are non-vanishing, is $\Pi^i_{0I}=\Pi^i_{IJ}y^J$ for an arbitrary $\mathbb{R}^3$-valued scalar $y$. After ``solving'' the remaining constraints~(\ref{simpl2}) one arrives at a reduced phase space parametrised by an $SU(2)$ connection and the new f\/ield $y$ together with their conjugate momenta. This system is free of second class constraints, but the f\/irst class constraints, in particular the Hamiltonian constraint (\ref{konstf}), take a rather complicated form in the new variables. One arrives at the usual simpler formulation for LQG by further reducing to the submanifold $y^J\equiv 0$. In terms of the original variables, $\Pi^i_{0I}$ is then constrained to vanish, and the dynamical variables are the $SU(2)$ connection $A^I_i\equiv\half{\epsilon^I}_{JK}A^{JK}_i$ and its conjugate momentum, the triad $E^i_I$. Equivalently, this constraint formulation can be obtained by imposing {\it time gauge} $E^0\equiv 0$ in the action (\ref{hamact}). Since the local gauge group is now compact, a Yang--Mills-type quantisation is possible in LQG.

In the following section, we will see that the addition of a $\gamma$-dependent term to the Hamiltonian constraint is very natural from the viewpoint of the geometry of the space of connections.

\section{Geometry of the space of connections}
\label{geomconn}

In this section, we recall the ADM formulation \cite{adm} of general relativity in metric variables whose Hamiltonian constraint def\/ines a metric on Wheeler's superspace of 3-metrics, and extend the discussion to the space of connections. For the superspace of metrics we essentially follow the review \cite{giulinisuper} where many more details on the geometry and topology of this space can be found.

The relevant quantities in the metric formulation, def\/ined on a spacelike surface $\Sigma$, are a~Riemannian metric $h$ and a~symmetric tensor $K$ which becomes the extrinsic curvature of $\Sigma$ in the full spacetime. The Hamiltonian constraint for vacuum GR has the form \cite{giulinisuper}
\begin{gather}
||K||^2_h - (\Tr_h (K))^2 - R(h) = 0,
\label{metricconst}
\end{gather}
where indices are raised and lowered with $h$.

Since $K$ is essentially conjugate to the metric $h$, one can take the part of (\ref{metricconst}) quadratic in $K$ to def\/ine an ``ultralocal'' (i.e.\ only involving multiplication of functions at the same point in $\Sigma$ without spatial derivatives)  metric on the space of 3-metrics. More generally, one could consider a family of metrics
\begin{gather}
G_{\alpha(x),\lambda}(k,l)=\int_{\Sigma}d^3x\,\alpha \sqrt{h}\big(h^{ij}h^{mn}k_{im}l_{jn}-\lambda(h^{ij}k_{ij})(h^{mn}l_{mn})\big) ,
\label{wheelermetric}
\end{gather}
invariant under dif\/feomorphisms in $\Sigma$, where $k$ and $l$ are elements of the tangent space to $h$ in the space of Riemannian metrics, and $\alpha$ is some positive function that may be set to one.

The constraint (\ref{metricconst}) picks the value $\lambda=1$, and it can be shown that this value is special from purely geometric considerations of the action of dif\/feomorphisms on the space of metrics~\cite{giulinisuper}.

Instead of a metric on the space of metrics, one often considers the associated bilinear form $\Tr(k\cdot l)-\lambda\Tr(k)\Tr(l)$ on the space of $(3\times 3)$ symmetric matrices. One can write the metric as $h=E^T E$, where $E$ is the three-dimensional frame f\/ield which lies in the coset space $GL(3)/O(3) \simeq \mathbb{R}\times SL(3)/SO(3)$; there is a one-parameter family of metrics
\begin{gather*}
ds^2 = \Tr\big(E^{-1}dE\,E^{-1}dE\big)+\beta\big(\Tr\big(E^{-1}dE\big)\big)^2
\end{gather*}
on this coset. The two terms can now be identif\/ied as the Killing metric on $\mathfrak{sl}(3)$ and the trivial measure on $\mathbb{R}$, respectively.

Now consider the space $\A$ of all connections in $\Sigma$ def\/ined above. This is a vector space with a right action by the group $\mathfrak{G}$ of gauge transformations; inf\/initesimally, a gauge transformation is a $\mathfrak{g}$-valued function $T(x)$ on $\Sigma$ which induces a vector f\/ield on $\mathfrak{A}$, namely
\begin{gather*}
V_{T(x)}=\int_{\Sigma} d^3 x\, \big(D_i T^{ab}(x)\big)\frac{\delta}{\delta\Omega_i^{ab}(x)} \equiv \int_{\Sigma} d^3 x \,\big(d_i T^{ab}(x)+[\Omega_i,T]^{ab}(x)\big)\frac{\delta}{\delta\Omega_i^{ab}(x)},
\end{gather*}
such that $T\mapsto V_T$ is a Lie homomorphism, i.e.\ $V_{[T,T']}=[V_T,V_{T'}]$. There is also an action by the dif\/feomorphism group of $\Sigma$ on $\mathfrak{A}$, given by the Lie derivative (analogous to \cite[Section~3]{giulinisuper}).

The Hamiltonian constraint in connection variables (\ref{konstf}) def\/ines a diagonal (but not ultralocal) (inverse) metric on space $\mathfrak{A}$, which in the cotangent space to $\A$ at $\Omega$ is given by
\begin{gather}
G_{\alpha(x),\gamma}(\pi,\tau) = \int_{\Sigma}d^3 x \,\alpha(x){\pi^{ia}}_c(x)\tau^{jbc}(x)\left(R_{ijab}[\Omega](x)- \frac{s}{2\gamma}\epsilon_{abef}R^{ef}_{ij}[\Omega](x) \right).
\label{connmetric}
\end{gather}

If $\alpha(x)$ is a scalar function, we integrate a scalar density of weight two in $\Sigma$ and the metric is not invariant under dif\/feomorphisms in $\Sigma$; one could instead use a scalar density of weight minus one, such as $\big(\det(\vec{R}_{12},\vec{R}_{13},\vec{R}_{23})\big)^{-1/2}$, which is only def\/ined where this determinant is non-zero, to construct a scalar under 3-dif\/feomorphisms. The bilinear form associated to (\ref{connmetric}) is
\begin{gather}
(G_{\gamma})^{[ac][bd]}_{ij}=\left[\eta^{cd}\left(R^{ab}_{ij}- \frac{s}{2\gamma}{\epsilon^{ab}}_{ef}R^{ef}_{ij}\right)-\eta^{ad}\left(R^{cb}_{ij}- \frac{s}{2\gamma}{\epsilon^{cb}}_{ef}R^{ef}_{ij}\right)\right]-(b\leftrightarrow d)\,,
\label{bilform}
\end{gather}
viewed as an ($18\times 18$) matrix where ${}^{[ac]}_i$ label rows and ${}^{[bd]}_j$ label columns. The Barbero--Immirzi parameter $\gamma$ appears to be the analogue of the parameter $\lambda$ appearing in (\ref{wheelermetric}) in the case of 3-metrics, in the following sense: If we consider the action of $G$ gauge transformations on the frame f\/ield, $E^b_i \rightarrow{\Lambda^b}_c E^c_i$, the momenta $\pi^i_{ab}$ transform in the adjoint representation, as does the curvature 2-form $R^{ab}_{ij}$. Hence, for a form linear in the curvature, we need a three-index invariant tensor for the adjoint representation of $G$ to def\/ine an invariant metric. For simple Lie algebras, such a tensor is given by the structure constants of the Lie algebra contracted with the Killing form. The algebras $\mathfrak{so}(3,1)$ and $\mathfrak{so}(4)$ are semisimple and we have the situation described by Wise \cite[Appendix~B]{wise2} for the case of symmetric bilinear forms: There is a family of invariant trilinear forms, antisymmetric in the f\/irst two arguments, given by
\begin{gather}
\nu(X,Y,Z)=\kappa([X,Y],(c_0 + c_1\star) Z) ,\qquad X,Y,Z\in\mathfrak{g},
\label{trilinearforms}
\end{gather}
(where $\kappa$ is the Killing form), and these span all such forms, since the common complexif\/ication of $\mathfrak{so}(3,1)$ and $\mathfrak{so}(4)$ splits as $\mathfrak{so}(4,\mathbb{C})\simeq \mathfrak{sl}(2,\mathbb{C})\oplus \mathfrak{sl}(2,\mathbb{C})$, and for each of the factors all invariant trilinear forms have the form $\kappa([X,Y],Z)$~\cite{izquierdo}. Hence, just as in the case of Wheeler's superspace, one has two possible terms in the metric. Their relative weight is characterised by $\lambda$ in the metric case and by $\gamma$ in the connection case.

For $G=SO(3,1)$, it is easy to check that for $\gamma=\pm {\rm i}$, and only for these values, the bilinear form (\ref{bilform}) is degenerate independent of the connection $\Omega$, and hence these values are special. This is because for $\gamma=\pm {\rm i}$ (\ref{bilform}) involves a projection of the complexif\/ication $\mathfrak{so}(4,\mathbb{C})$ on its \mbox{(anti-)}self-dual part. If $G=SO(4)$, an identical calculation gives a degenerate bilinear form~(\ref{bilform}) if $\gamma=\pm 1$, again corresponding to a projection on the (anti-)self-dual part of $\mathfrak{so}(4)$. Of course $\gamma$ is normally taken to be real; nevertheless this observation is a re-statement of the fact~\cite{samuel} that in the Barbero formulation of GR the connection cannot be interpreted as a space-time connection unless $\gamma$ assumes one of the values $\pm {\rm i}$ used in the original Ashtekar formulation~\cite{ashtekar}.

In this sense, the values $\gamma=\pm {\rm i}$ or $\gamma=\pm 1$ are analogous to the preferred value $\lambda=1$ in the metric formulation. On the other hand, this argument does not give any preferred {\it real} values of $\gamma$ for $G=SO(3,1)$. In particular, the limit $\gamma\rightarrow\infty$ is not special, and from this viewpoint it seems more natural to allow general $\gamma$.

In the general case (for real $\gamma$) the bilinear form is denegerate if $R^{ab}- \frac{s}{2\gamma}{\epsilon^{ab}}_{cd} R^{cd}$ only takes values in a proper subspace of $\mathfrak{g}$, or equivalently if $R^{ab}$ only takes values in a subalgebra $\mathfrak{a}$ that  satisf\/ies ${\rm span}\{\mathfrak{a}\cup \star(\mathfrak{a})\}\neq\mathfrak{g}$, where $\star$ is the Hodge dual in the Lie algebra as in (\ref{hodge}). The subalgebras $\mathfrak{a}$ leading to a degenerate bilinear form (in the case $G=SO(3,1)$) are obviously all one- and two-dimensional subalgebras of $\mathfrak{g}$, as well as the Bianchi algebras V, ${\rm VII}_0$ and ${\rm VII}_a$ and the four-dimensional algebra $\mathfrak{sim}(2)$\footnote{For an overview of applications of this maximal subgroup of the Lorentz group, see~\cite{garytalk}.}.

If the curvature only takes values in a subalgebra of $\mathfrak{g}$ (everywhere), by the Ambrose--Singer theorem the holonomy of the connection is a subgroup of $G$, and so the connection is reducible. At these connections the action of $\mathfrak{G}$ has f\/ixed points. If one wanted the coset $\mathfrak{A}/\mathfrak{G}$, for compact~$G$, to be a smooth manifold, these connections would have to be excluded, or the allowed gauge transformations modif\/ied accordingly \cite{singer}\footnote{Compare the analogous discussion for the space of Riemannian metrics in \cite{giulinisuper}.}. The metric obtained from the Hamiltonian constraint hence f\/its in with the requirement that the action of $\mathfrak{G}$ be free.

One is tempted to stress here that the number of four spacetime dimensions is very special; in any other number of dimensions the rotation group is simple and the choice of invariant tensor for the adjoint representation would be unique. Let us brief\/ly contrast the previous discussion with the simpler case of 3 dimensions; here the relevant GR action is (the Lorentz group here is $G=SO(2,1)$ or $G=SO(3)$)
\begin{gather*}
S = \frac{1}{16\pi G}\int_{\Sigma\times\mathbb{R}}\! \epsilon_{abc} e^a\wedge R^{bc}[\omega] = \frac{1}{16\pi G}\int \! dt\int_{\Sigma}\!\epsilon_{abc}\big({-}\dot{\Omega}^{ab}\wedge E^c+\chi^a R^{bc}[\Omega]-\Xi^{ab} D^{(\Omega)}E^c\big) .
\end{gather*}
As in the case of four dimensions it is convenient to treat $\chi^a$ and $\Xi^{ab}$ as Lagrange multipliers. There is also again a set of primary constraints determining the momenta conjugate to the six connection components $\Omega^{bc}_j$ as functions of the variables $E^c_i$,
\begin{gather*}
\pi^i_{ab}=-\frac{1}{16\pi G}\epsilon^{ij}\epsilon_{abc}E^c_j ;
\end{gather*}
these can be inverted so that one can express $E^a_i$ directly in terms of $\pi^i_{ab}$ and formulate the theory in terms of the variables $(\Omega^{ab}_i,\pi^i_{ab})$ by the 3+3 constraints
\begin{gather*}
\mathcal{C}_a \equiv \epsilon_{abc}\epsilon^{ij}R^{bc}_{ij} ,\qquad \mathcal{G}_{ab} \equiv D_i^{(\Omega)}\pi^i_{ab} .
\end{gather*}
The theory has no local degrees of freedom. As suggested in \cite{thiemreduced}, if the nondegeneracy condition $\det\big(\pi^i_{ab}\pi^{ab}_j\big)\neq 0$ is satisf\/ied, one can construct a scalar constraint $\mathcal{H}=\epsilon_{ij}\epsilon^{efg}{\pi^{ia}}_e \pi^j_{fg}\mathcal{C}_a$ which again def\/ines a (degenerate) metric on the space of connections. In ``dualised'' notation $\pi^i_a=\half {\epsilon_a}^{bc}\pi^i_{bc}$, this part is proportional to $\pi^i_a\pi^j_b R^c_{ij}{\epsilon^{ab}}_c$, and so corresponds to the trilinear form on the adjoint representation of $G$ constructed from the structure constants. Here $\mathfrak{so}(3,\mathbb{C})\simeq \mathfrak{sl}(2,\mathbb{C})$ and there is no possibility to add a second term to the constraint.

\section{GR from a geodesic principle?}

The Hamiltonian constraint (\ref{konstf}) can be understood as a geodesic principle in the space of connections, as detailed in \cite{smolinsoo}; it follows from the simple action
\begin{gather}
S = \int dt\int_{\Sigma} d^3 x\,\frac{1}{2N}\dot{\Omega}^{ab}_i \dot{\Omega}^{cd}_j\,(G_{\gamma})_{abcd}^{ij}[\Omega] ,
\label{oakschn}
\end{gather}
or in canonical form,
\begin{gather*}
S = \int dt\int_{\Sigma} d^3 x\,\left(\dot{\Omega}^{ab}_i\pi^i_{ab}-\frac{N}{2}\pi^i_{ab} \pi^j_{cd} (G_{\gamma})^{abcd}_{ij}[\Omega]\right) ,
\end{gather*}
so that the Hamiltonian constraint~(\ref{konstf}) follows; the other constraints are then required by consistency of $\mathcal{H}\equiv 0$ under time evolution, as shown in~\cite{barros}. The analogy of~GR to a relativistic particle in a certain background metric has been used by many authors to discuss conceptual issues in quantum gravity, such as the properties of dif\/ferent two-point functions def\/ining an inner product~\cite{hallort}.

In (\ref{oakschn}), $(G_{\gamma})_{abcd}^{ij}$ is the inverse of the bilinear form (\ref{bilform}), which we recall is
\begin{gather*}
(G_{\gamma})^{[ac][bd]}_{ij}=\left[\eta^{cd}\left(R^{ab}_{ij}- \frac{s}{2\gamma}{\epsilon^{ab}}_{ef}R^{ef}_{ij}\right)-\eta^{ad}\left(R^{cb}_{ij}- \frac{s}{2\gamma}{\epsilon^{cb}}_{ef}R^{ef}_{ij}\right)\right]-(b\leftrightarrow d) ,
\end{gather*}
and is invertible for generic $\gamma$, unless the curvature $R^{ab}$ of $\omega^{ab}$ only takes values in certain subalgebras of $\mathfrak{g}$. Such connections have to be excluded from our conf\/iguration space for (\ref{oakschn}) to be def\/ined, and general relativity to be understood as free fall on superspace. An interesting question that we cannot answer here would be whether geodesics can always be extended to inf\/inite parameter length, i.e.\ whether one will encounter degenerate connections following a~geodesic.

In the metric case, the constraint (\ref{metricconst}) is not just quadratic in the canonical momenta, but involves, even for pure gravity, a ``potential term'' depending on the scalar curvature $R(h)$. However, as shown by Greensite \cite{greensite}, the constraint (\ref{metricconst}) for gravity, possibly coupled to scalar and vector f\/ields and with cosmological constant, can be rewritten as a geodesic equation on an extended ``superspace'', the conf\/iguration space of the theory, for an appropriate metric. The proof rests on the following steps:
\begin{itemize}\itemsep=0pt
\item As in (\ref{oakschn}), one only needs to consider the Hamiltonian (scalar) constraint, since the other constraints are generated by consistency (conservation of $\mathcal{H}=0$ in time). One can make specif\/ic gauge choices restricting the lapse and shift functions.
\item One only assumes a Hamiltonian constraint of the general form
\begin{gather}
\mathcal{H}=G^{ab}(q)p_a p_b+U(q),
\label{greensiteform}
\end{gather}
where $q^a$ are coordinates on the conf\/iguration space and $p_a$ are conjugate momenta, $G^{ab}$ is a non-degenerate symmetric form (so that one can solve for the momenta $p_a$ in terms of velocities $\dot{q}^a$), and $U(q)$ an arbitrary ``potential''. Then one shows that geodesic particle motion with respect to a metric $g_{\mu\nu}$ is equivalent to motion in a potential $\phi$ for a metric~$G_{\mu\nu}$ where $g_{\mu\nu}=\phi G_{\mu\nu}$. (\ref{greensiteform}) can therefore be viewed as the geodesic equation for an appropriate supermetric.
\end{itemize}
In particular, no explicit form of $G^{ab}$ is needed to prove the result. Trying to apply a similar reasoning to GR in connection variables, we immediately notice that matter f\/ields as well as the cosmological constant would couple to momenta (namely the triad) and hence would lead to a momentum-dependent ``potential'' in the Hamiltonian constraint. The statement that GR is free fall in the space of connections only holds for pure gravity without cosmological constant.

\section{Symmetries and the need for third quantisation}

Loop quantum gravity is the canonical quantisation of general relativity in connection variables; quantum states can be thought of as functionals $\psi[A]$ on the space of generalised connections~$\overline{\mathfrak{A}}$ on which an action of a certain set of operators corresponding to phase space functions is def\/ined. In this sense, one has performed a {\it first quantisation} analogous to single-particle quantum mechanics. When this quantisation procedure is applied to the relativistic particle, with mass-shell constraint $\mathcal{C}\equiv g^{\mu\nu}p_{\mu}p_{\nu}+m^2$, one faces the well-known dif\/f\/iculties of def\/ining a~positive-def\/inite inner product and corresponding probability interpretation, which are overcome by splitting the solutions to the constraint into positive- and negative-frequency subspaces. In the second quantised theory, the dif\/ferent sectors are then identif\/ied with operators creating and annihilating particles. However, quantum f\/ield theory is meaningful also when a symmetry allowing the def\/inition of positive and negative frequency is absent and there is no unambiguous particle concept.

This line of reasoning was pushed further in a paper by Kucha\v{r} \cite{kuchar}; it was argued that in order to have a meaningful ``one-particle'' (i.e.\ one-universe) interpretation of quantum geo\-metrodynamics, one would require a symmetry on superspace analogous to time translation on Minkowski space. More concretely, one requires a {\it conditional symmetry}: This would be the existence of a quantity linear in momenta,
\begin{gather*}
K=\int_{\Sigma} d^3 x\,\kappa_{ij}(x)[h]p^{ij}(x) ,
\end{gather*}
such that on the constraint surface where $\mathcal{H}$ and $\vec{\mathcal{H}}$ vanish one has
\begin{gather*}
\big[K,\mathcal{H}[N]+\vec{\mathcal{H}}[\vec{N}]\big]\approx 0\qquad\forall\,N(x),\vec{N}(x) ,
\end{gather*}
where $\mathcal{H}[N]$ and $\vec{\mathcal{H}}[\vec{N}]$ are the usual smeared forms of the constraints, $\mathcal{H}[N]=\int d^3 x\,N(x)\mathcal{H}(x)$. Kucha\v{r} proceeded to show that no such quantity exists, and argued that there is therefore no meaningful one-universe theory of quantum geometrodynamics, so that one has to go to a ``third quantised'' theory where f\/ixed ``universe number'' is no longer the central concept. The idea of third quantisation was then taken up by other authors, e.g.\ in \cite{giddstrom}.

For GR in connection variables, a modern third quantisation approach that grew out of loop quantum gravity is group f\/ield theory (GFT) \cite{gftreview}, which is however not formulated on the (inf\/inite-dimensional) space of connections on a continuous manifold $\Sigma$, but on the group manifold $G^n$, interpreted as the space of holonomies describing parallel transport along $n$ given paths, thereby encoding the geometry of a discrete structure such as a $d$-simplex. A more direct ``second quantisation'' of loop quantum gravity would be a tentative formalism for third quantisation based on {\it continuous} connections, which could be an appropriate description for a~continuum limit of GFT~\cite{3rdquant}.

Without going further into the details of the formalism outlined in \cite{3rdquant}, we contrast the situation for the space of metrics with the proposal in \cite{smolinsoo} for a time variable on the space of connections for $\gamma=\pm \sqrt{s}$ so that the connection is self-dual and can be thought of as a one-form valued in $\mathfrak{su}(2)\otimes\bR$ (Euclidean) or $\mathfrak{su}(2)\otimes\bC$ (Lorentzian); the gauge group is then reduced to $G=SU(2)$ and the Hamiltonian constraint becomes \cite{rovelli}\footnote{Here we denote Lie algebra indices by $a,b,\ldots$ although the gauge group is supposedly $SU(2)$ since in the Lorentzian theory the connection is really valued in $\mathfrak{sl}(2,\bC)$, the full Lorentz group.}
\begin{gather*}
\mathcal{H}\equiv E^i_a E^j_b F_{ij}^{ab}[A] ,
\end{gather*}
where $E^i_a$ are conjugate to the $\mathfrak{su}(2)$ connection $A_i^a$.

In \cite{smolinsoo} several arguments are given for the interpretation of the Chern--Simons \cite{chernsim} invariant
\begin{gather}
T[A] = \int_\Sigma \Tr\left(A\wedge dA+\frac{2}{3}A\wedge A\wedge A\right)
\label{cherns}
\end{gather}
as a natural time variable on the space of connections. Since the variation of (\ref{cherns}) is $\delta T\sim \int \delta A^a\wedge \epsilon_{abc}F^{bc}$, the momentum conjugate to $T$ is (up to a constant)
\[
P_T = \left(\int_\Sigma d^3 x\,\sqrt{\det F}\right)^{-1}\int_\Sigma d^3 x\,\sqrt{\det F} F^a_i E_a^i ,
\]
where $F^i_a=\frac{1}{4}\epsilon_{abc}\epsilon^{ijk}F_{jk}^{bc}$ are interpreted as elements of a matrix $F$ and $F^a_i$ are the elements of the inverse matrix $F^{-1}$, $F^i_a F^a_j=\delta^i_j$. Note that in this formalism $\vec{E}_a$ and $\vec{F}_a$ are vector densities and $(\det F)$ has density weight two so that $P_T$ is independent of the choice of coordinates on~$\Sigma$. If $T$ is accepted as a time variable and $P_T$ is accepted as the analogue of frequency, one could now proceed to split the Hilbert space of loop quantum gravity into positive- and negative frequency components; no such proposal seems to have been made in the existing literature. The Chern--Simons invariant does of course feature in the def\/inition of the Kodama state \cite{kodama,smolinsoo}
\[
\psi_{{\rm Kod}}[A]=\exp\left(\frac{3}{16\pi\Lambda} T[A]\right) ,
\]
claimed to be an exact solution in quantum gravity corresponding to de Sitter space, whose signif\/icance in the full theory is however disputed \cite{wittenkodama}. Note that the analogous time variable for the metric formalism of GR, proposed by York~\cite{york}, is the trace of the extrinsic curvature which depends on both ``positions'' and ``momenta'', unlike~(\ref{cherns}) for connections.

\section{Summary}
In this short review paper, we have taken the view that a careful study of the space of connections, in light of its role as the conf\/iguration space for general relativity, is a f\/irst important step towards understanding gravity in connection variables, f\/irst at the classical level and then for its quantisation in LQG. We have focussed on its possible metric structure, an interpretation of dynamics as free fall in such a metric, and the existence of a time variable which is important for a physical interpretation of the quantum theory. When reviewing known properties of the space of connections, we compared them with those of the traditional superspace of geometrodynamics.

The Hamiltonian constraint of GR def\/ines a metric on the space of connections whose integrand in its general form, with the Barbero--Immirzi parameter $\gamma$ appearing as a free parameter giving the relative weight of two possible terms, can be viewed as the most general invariant three-index tensor for the adjoint representation of the gauge group $G$ in which both the curvature of the connection and the conjugate momenta live. The existence of two possible tensors lies in the fact that $G$ is not simple in four dimensions. In contrast to the metric case, the Hamiltonian constraint GR is quadratic in momenta with no ``potential term'' present. Therefore, on the one hand it can directly be viewed as def\/ining a geodesic principle on the space of connections, while on the other hand the addition of matter or a cosmological constant would not lead to a ``potential'' term, so that a geodesic interpretation can only be given to pure GR. Finally we discussed the need for a quantity on the conf\/iguration space of the theory that plays the role of energy for the quantum mechanics of a single particle. For the space of connections, the Chern--Simons invariant might be a natural candidate, but its signif\/icance in the interpretation of the Hilbert space of LQG remains at present rather unclear and deserves further study.

\subsection*{Acknowledgements}
I should like to thank Daniele Oriti for initiating this study of the geometry of the space of connections and for fruitful discussions, and an anynomous referee for comments that led to an improvement of presentation, particularly concerning Hamiltonian dynamics.

\pdfbookmark[1]{References}{ref}
\LastPageEnding

\end{document}